\documentclass[prb,a4paper,twocolumn,groupedaddress,english,showpacs]{revtex4}
\usepackage{babel}
\usepackage{xcolor}
\usepackage{graphicx}
\usepackage{amsmath}
\usepackage{amssymb}

\definecolor{darkblue}{rgb}{0.1,0.2,0.6}
\definecolor{darkred}{rgb}{0.8,0.1,0.2}
\usepackage[colorlinks,citecolor=darkblue,linkcolor=darkred,urlcolor=darkblue]{hyperref} 
\usepackage[all]{hypcap} 

\makeatletter
\adddialect\l@English\l@english
\makeatother

\newcommand{\bra}[1]{\langle\,#1\,|}
\newcommand{\ket}[1]{|\,#1\,\rangle}
\newcommand{\bracket}[1]{\langle\,#1\,\rangle}

\newcommand{\E}{\mathrm{e}}
\newcommand{\D}{\mathrm{d}}

\newcommand{\cf}{\textit{cf.} }
\newcommand{\ie}{\textit{i.e.} }
\newcommand{\eg}{\textit{e.g.} }
\newcommand{\vs}{\textit{vs.} }
\newcommand{\etal}{\textit{et al.} }

\newcommand{\tr}{\mathrm{Tr}}

\definecolor{commentcolor}{rgb}{0.1,0.2,0.6}
\definecolor{commentcolor2}{rgb}{1,0,0}
\definecolor{commentcolorF}{rgb}{0,0,1}
\definecolor{commentcolorD}{rgb}{1,0.1,.1}
\definecolor{todocolor}{rgb}{0.8,0.1,0.2}

\begin{document}
\title{Improving entanglement and thermodynamic R\'enyi entropy measurements in quantum Monte Carlo}

\author{David J. Luitz}
\affiliation{Laboratoire de Physique Th\'eorique, IRSAMC, Universit\'e de Toulouse,
{CNRS, 31062 Toulouse, France}}
\email{luitz@irsamc.ups-tlse.fr}
\author{Xavier Plat}
\affiliation{Laboratoire de Physique Th\'eorique, IRSAMC, Universit\'e de Toulouse,
{CNRS, 31062 Toulouse, France}}
\author{Nicolas Laflorencie}
\affiliation{Laboratoire de Physique Th\'eorique, IRSAMC, Universit\'e de Toulouse,
{CNRS, 31062 Toulouse, France}}
\author{Fabien Alet}
\affiliation{Laboratoire de Physique Th\'eorique, IRSAMC, Universit\'e de Toulouse,
{CNRS, 31062 Toulouse, France}}
\date{\today}

\begin{abstract}
    We present a method for improving measurements of the entanglement R\'enyi entropies in quantum
    Monte Carlo simulations by relating them with measurements of participation R\'enyi entropies.
    Exploiting the capability of building improved estimators for the latter allows to obtain very good estimates for entanglement R\'enyi entropies. When considering a full system instead of a bipartition, the method can be further ameliorated providing access to the thermodynamic R\'enyi entropies with high accuracy.
    We also explore a recently-proposed method for the reconstruction of the entanglement
    spectrum from entanglement R\'enyi entropies and finally show how
    potential entanglement Hamiltonians may be tested for their validity using a comparison with
    thermal R\'enyi entropies.
\end{abstract}
\maketitle
\section{Introduction}
Quantum entanglement has been of interest since the early days of quantum mechanics~\cite{Einstein35}. The quantification of the entanglement in interacting many body quantum systems has attracted a lot of attention during the last decade for several fundamental and practical reasons~\cite{Calabrese_Special_JPA09}.
Entanglement properties of one-dimensional quantum problems can be treated fully analytically only
in a limited number of cases (see {\it e.g.}
Refs.~\onlinecite{jin_quantum_2004,fan_entanglement_2004}), or asymptotically for conformally
invariant~\cite{calabrese_entanglement_2004} or some disordered systems~\cite{Refael04}. They remain
generically accessible to numerical Density Matrix Renormalization Group (DMRG) calculations,
provided the entanglement between the subsystems is not too large~\cite{Schollwock05}. For higher dimensional systems however, exact methods are much more difficult to implement. Nevertheless, remarkable progresses have been made recently, {\it{e.g.}} using series expansions~\cite{kallin_anomalies_2011,singh_thermodynamic_2012}, numerical linked cluster expansion~\cite{kallin_entanglement_2013,kallin_corner_2014}, or using quantum Monte Carlo (QMC) simulations~\cite{Alet07,Kallin09,hastings_measuring_2010,melko_finite-size_2010,humeniuk_quantum_2012,Grover13,Assaad14, chung_entanglement_2013,Helmes14,Herdman14}, which is precisely the topic of the present work.

  In nonfrustrated quantum spin systems, standard thermodynamic observables can be obtained to
very high accuracy within QMC simulations~\cite{sandvik_computational_2010}. Here we are interested in the R\'enyi entanglement entropy (EE) 
\begin{equation}
    S_{A,q}^{\text{E}} = \frac{1}{1-q} \ln \rho_A^q,
    \label{eq:EE}
\end{equation}
where $\rho_A$ is the reduced density matrix, assuming that $A$ is a subsystem imbedded in a larger system.
Clearly $S_{A,q}^{\text{E}}$ cannot be related to a simple thermodynamic observable, {\it{e.g.}} a correlation function (except for non-interacting systems~\cite{Chung01,Audenaert02}). At zero temperature, Hastings \etal
\cite{hastings_measuring_2010} developed a technique based on the introduction of a
``swap''-operator in a projector Monte Carlo approach to tackle this issue. At finite
temperature, several techniques have been explored, including temperature integration 
\cite{melko_finite-size_2010} and Wang Landau sampling \cite{inglis_wang-landau_2013}.

Perhaps the
most elegant method was brought forward by Humeniuk and Roscilde
\cite{humeniuk_quantum_2012}. Their method for the 
calculation of entanglement R\'enyi entropies of order $q$ for a subsystem $A$ in path integral
QMC methods is based on the observation~\cite{calabrese_entanglement_2004} that they are related to the ratio of 
partition functions $\mathcal{Z}_{A,q}/\mathcal{Z}_{\varnothing,q}$. Here,
$\mathcal{Z}_{A,q}$ is the partition function of $q$ replicas glued together at one
imaginary time slice on the subsystem $A$ only. $\mathcal{Z}_{\varnothing,q} = \mathcal{Z}^q$ is the
partition function of $q$ independent replicas. Here and from now on, $q$ is an integer $\ge 2$.

In a simulation which samples both partition functions in a generalized ensemble, proposing moves
between the two ensembles, the estimator for the entanglement R\'enyi entropy is given by\cite{humeniuk_quantum_2012}:
\begin{equation}
    \label{eq:EE_QMC}
    \bracket{S_q^{E}}_\text{MC} = \frac{1}{1-q} \ln \bracket{ \frac{N_A}{N_{\varnothing}} }_\text{MC},
\end{equation}
where $N_A$ is the number of QMC configurations observed in the glued ensemble, while
$N_\varnothing$ is the number of QMC configurations seen in the independent ensemble. 
This method becomes inefficient for too large entropies, which lead to very small $N_A$
and large $N_\varnothing$. This is a problem of rare events which is also known in the
related context of participation R\'enyi (PR) entropies\cite{luitz_universal_2014, luitz_shannon_2014,
luitz_participation_2014} and prohibits the estimation of large entropies in finite simulation time. The problem can however be cured by the application of the ``ratio trick''~\cite{deForcrand_2001,hastings_measuring_2010}, calculating
the entanglement entropy by a stepwise increase of the subsystem $A$.

Let us give a description of the results presented in this article, along with its organization. We will first start from equation \eqref{eq:EE_QMC} to show how the measurement of the
entanglement entropy is related to the basis dependent participation R\'enyi entropies $S_q^\text{PR}$~\cite{stephan_shannon_2009,luitz_participation_2014} (also called Shannon-R\'enyi entropies in the litterature). The
basic idea is to split the extended ensemble in its two parts and to simulate the ensemble of
independent replicas and the ensemble of replicas that are glued together on subsystem $A$ (see
Fig. \ref{fig:glued}) separately. In Sec.~\ref{sec:RC}, we derive the following relation:
\begin{equation}
    S_q^{\text{E}} = S_q^{\text{PR}} - C_q^{\text{R}}.
    \label{eq:EEfromPR}
\end{equation}
which relates the entanglement entropy $S_q^{\text{E}}$ to the difference between participation R\'enyi entropy $S_q^{\text{PR}}$ and the \emph{replica correlation} $C_{q}^\text{R}$, which is introduced in Sec.~\ref{sec:RC} and is defined in the glued
  ensemble. 
  
  Remarkably, this combination of two basis-dependent quantities will hint towards a more efficient calculation of the entanglement entropy, for two different reasons. First, large entanglement entropies can be obtained (both in our setup and the one used in Ref.
  \onlinecite{humeniuk_quantum_2012}) when using a QMC computational basis where the participation entropies  $S_q^{\text{PR}}$ are small. Second, we will introduce in Sec.~\ref{sec:PR} several improved Monte Carlo estimators which will greatly increase the precision on $S_q^{\text{PR}}$ and consequently on $S_q^{\text{E}}$.

As an interesting by-product, our scheme allows to compute the thermodynamic R\'enyi entropy $S_q^{\text{th}}$ in the specific case where the full system $\overline{\varnothing}$ and the subsystem $A$ are identical, as developed in Sec.~\ref{sec:renyi}. We will show there how the
translation invariance in imaginary time can also be used to construct an improved estimator for the
replica correlation $C_q^{\text{R}}$ , leading to a very accurate result for the thermodynamic
R\'enyi entropy. Let us emphasize that the thermodynamic R\'enyi entropy (for integer $q>1$) can be
calculated from two \emph{standard} QMC simulations (using
independent replicas) \emph{without the need of implementing a different stochastic process or
Wang-Landau sampling}.

The new methods are extensively tested and their efficiency discussed in Sec.~\ref{sec:results}
where we provide several results on quantum spin chains and ladders. We also explore in this section
the possibility to reconstruct the entanglement spectrum from R\'enyi entanglement
entropies~\cite{song_bipartite_2012,chung_entanglement_2013},  discussing the limitations of this
approach. There, we also propose an alternative method to test a putative entanglement Hamiltonian,
based on the comparison between R\'enyi entanglement entropies and the R\'enyi thermodynamic
entropies of the entanglement Hamiltonian. 

Finally, Sec.~\ref{sec:conc} draws conclusions on our work while the appendices contain details on the improved estimator derivation as well as on using symmetry sectors when measuring entanglement entropies.

\section{Method}
\label{sec:RC}

    \begin{figure}
        \begin{center}
            \includegraphics[width=\columnwidth]{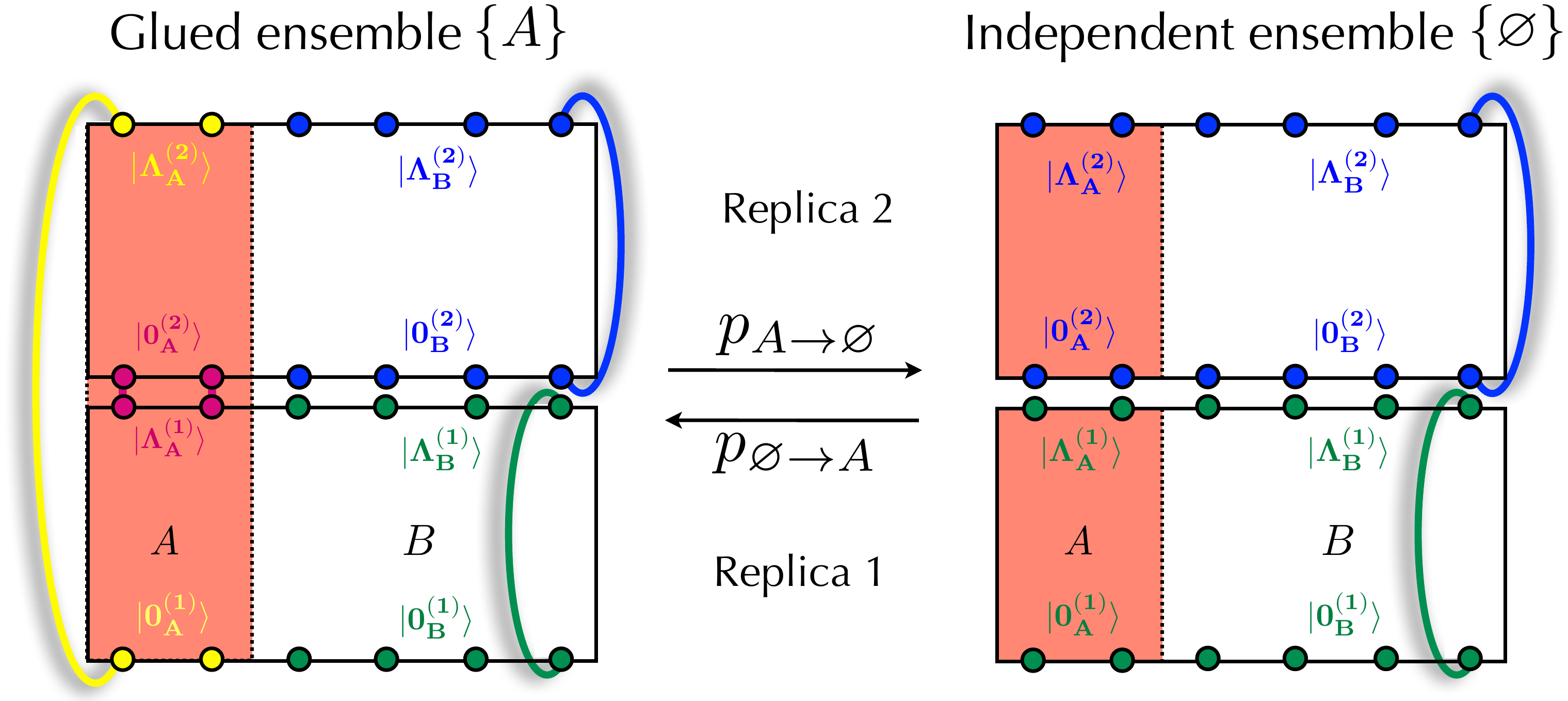}
        \end{center}
        \caption{\label{fig:glued} 
            (Color online) Example of a QMC configuration with $q=2$ replicas in the glued ensemble (left) and in the independent ensemble (right) with a notable difference of boundary conditions in imaginary time. Using the stochastic series expansion~\cite{sandvik_computational_2010} notations with operator strings of length $\Lambda$ for illustration, these boundary conditions are as follows. Left: in the \emph{glued ensemble}, the subsystem $A$ state $\ket{\Lambda_A^{(1)}}$ at
            expansion slice $\Lambda$ of replica $1$ must be equal to $\ket{0_A^{(2)}}$ and consequently,
            $\ket{\Lambda_A^{(2)}}=\ket{0_A^{(1)}}$, while the operator string is $\Lambda$ periodic in the subsystem B
            for each replica $\alpha\in\{1,2\}$ (i.e. $\ket{0_B^{(\alpha)}}=\ket{\Lambda_B^{(\alpha)}}$).
          Right:  the \emph{independent ensemble} has the same cyclicity on subsystem $B$ as in the glued ensemble
            but also shows the condition $\ket{0_A^{(\alpha)}}=\ket{\Lambda_A^{(\alpha)}}$ for all replicas on
            subsystem $A$. Therefore, all replicas are truly independent.
        }
    \end{figure}

  The method proposed by Humeniuk and Roscilde \cite{humeniuk_quantum_2012} uses an extended ensemble
simulation (see Fig.~\ref{fig:glued}), which dynamically moves between the glued ($\{A\}$) ensemble and the independent ensemble
($\{\varnothing\}$) and records the ratio of Monte Carlo steps performed in the glued \vs independent
ensembles. For the equilibrium (Monte Carlo-) time $t$ evolution of the probability $P_A(t)$ to be in ensemble $A$, the
Master equation 
\begin{equation}
    \frac{\D P_A(t)}{\D t} = P_{\varnothing}(t) p_{\varnothing \to A} - P_{A}(t) p_{A \to \varnothing}
\end{equation}
holds, where $p_{\varnothing \to A}$ is the probability of moving from the independent ensemble to
the glued ensemble and $p_{A\to \varnothing}$ is the probability of the inverse move. 
In an equilibrated Markov chain of QMC configurations, the probability of
finding the glued ensemble is time-independent, and we therefore obtain:
\begin{equation}
    \frac{P_A}{P_\varnothing} = \frac{p_{\varnothing\to A}}{p_{A\to\varnothing}}.
\end{equation} and with equation \eqref{eq:EE_QMC}
\begin{equation}
    S_q^\text{E} = \frac{1}{1-q} \ln \left( \frac{ p_{\varnothing \to A} }{ p_{A \to \varnothing} }
    \right). 
\end{equation}

  Instead of calculating this ratio of probabilities in a single QMC calculation, let us now
concentrate on an estimation of $p_{\varnothing \to A}$ and $p_{A \to \varnothing}$ in separate
calculations. 

\subsection{Probability of leaving the independent ensemble $p_{\varnothing\to A}$}

  If the simulation is in the independent ensemble, the condition for moving to the glued ensemble
is given by finding identical states on the subsystem $A$ in all $q$ replicas. This corresponds to
identical states $\ket{0_A^{(i)}}$ for all replicas $i\in [1,2,\dots,q]$ following the convention of
Fig.~\ref{fig:glued}.
It is important to
note that in principle due to the cyclicity of the trace the time slice where the replicas would be
sewed together does not matter. However, we will not actually perform the step of moving to the
glued ensemble here, but will just think about how probable it is. Clearly, we then have:
\begin{equation}
    p_{\varnothing \to A} = P_{\text{identical},q,A}.
\end{equation}
But the probability $P_{\text{identical},q,A}$ of finding identical states on the subsystem $A$ in
all $q$ replicas is just given by the participation R\'enyi entropies\cite{luitz_universal_2014} of subsystem $A$
\begin{equation}
    S_{A,q}^{\text{PR}} = \frac{1}{1-q} \ln P_{\text{identical},q,A}.
\end{equation}
We will present in Sec.~\ref{sec:PR} improved estimators for estimating efficiently the participation R\'enyi entropies.

\subsection{Probability of leaving the glued ensemble $p_{A\to \varnothing}$}

  When the simulation explores the glued ensemble $\{A\}$, the condition of moving to the independent
ensemble is given by having identical states on top and bottom of each replica individually, such
as to meet the condition of the trace. Therefore, $p_{A\to\varnothing}$ can be estimated by
performing a simulation in the glued ensemble $\{A\}$ and recording how often this condition is met,
relative to the total number of QMC steps. We have
\begin{equation}
    p_{A\to\varnothing}=
    P_{ \{A\} }\left(\ket{0_A^{(1)}}= \ket{\Lambda_A^{(1)}} \text{~and}  \dots 
    \ket{0_A^{(q)}}=\ket{\Lambda_A^{(q)}}\right). 
\end{equation}
This step turns out to be the bottleneck of the method in Ref.~\onlinecite{humeniuk_quantum_2012}, as this probability decays exponentially with the
number of degrees of freedom in the subsystem $A$ and thus the \emph{replica correlation}
\begin{equation}
    C_{A,q}^{\text{R}} =  \frac{1}{1-q} \ln p_{A\to \varnothing}
\end{equation}
exhibits a \emph{volume law}.  Note that the participation R\'enyi part suffers similar exponentially small probabilities for which we can however improve the estimate (see Sec.~\ref{sec:PR}).
This volume law is directly related to the problem of low acceptance
rates in the standard method \cite{humeniuk_quantum_2012}.
It should be noted that $C_{A,q}^{\text{R}}$ is not an entropy in the sense of equation
\eqref{eq:EE_QMC} as in general no density matrix can be found that provides
$C_{A,q}^\text{R}$ for all $q$. In particular, $C_{A,q}^\text{R}$ can grow with $q$,
which is not possible for R\'enyi entropies.

\subsection{Entanglement entropy}

Combining the two results, we therefore arrive at the previously announced Eq.~\eqref{eq:EEfromPR} for measuring entanglement entropies, which is restated for clarity using the spatial region index:\begin{equation}
    S_{A,q}^{\text{E}} = S_{A,q}^{\text{PR}} - C_{A,q}^{\text{R}}.
\end{equation}
This relation between the R\'enyi form of the participation and entanglement entropies provides
very interesting insights in the performance of any method based on the idea by Humeniuk and Roscilde
\cite{humeniuk_quantum_2012}. It has been established that PR entropies show a volume law in local
bases, \cite{stephan_shannon_2009,stephan_renyi_2010,stephan_phase_2011,luitz_universal_2014,luitz_shannon_2014,luitz_participation_2014}. 
However the coefficient of the volume term is non-universal and depends on the basis. 
As entanglement entropies
usually display an \emph{area law} for condensed-matter ground-states~\cite{eisert_colloquium_2010}, the replica correlation $C_{A,q}^{\text{R}}$ necessarily has
to exhibit the same volume law (and correspondingly, the probability to leave the glued ensemble decreases exponentially with the number of degrees of freedom in $A$).

The same reasoning goes for the basis dependence: $S_{A,q}^{\text{PR}}$ depends on the basis, while
$S_{A,q}^{\text{E}}$ \emph{does not}. Thus, $C_{A,q}^{\text{R}}$ needs to be a basis dependent quantity,
too. As the probabilities to be observed in the QMC calculations are potentially very small, it is
beneficial to choose the basis in which they assume {\it larger} values. 
Therefore, one should try to choose a computational basis in which the participation entropy $S_{A,q}^{\text{PR}}$
is the {\it smallest}. One notable example is a
simulation of the XX model $H_{\rm XX}=\sum_{\langle i,j \rangle} S^x_i S^x_j+S^y_iS^y_j$ in the basis in which $S_x$ is diagonal instead of the usual $S_z$ basis.

\section{Improved estimators}
\label{sec:PR}

Too large entropies lead in general to statistical issues in the QMC simulations. In order to tackle problems connected to the corresponding rare events, it is useful to
increase the number of Monte Carlo measurements as much as possible. Here, we present improved estimators that greatly enhance the precision of participation entropies $S_q^{\text{PR}}$ using all possible symmetries in imaginary time and real space. 

The starting point is the replica method introduced in Ref.~\onlinecite{luitz_universal_2014}. The basic idea is that in order to
measure the PR entropy $S_q^{\text{PR}}$ of a subsystem $A$ (which may coincide with the full
system), it is sufficient to estimate the probability of finding the same state $\ket{i}_A$ on the
part corresponding to subsystem $A$ of the state at operator string slice $i$. According to the
convention given in Fig.~\ref{fig:glued}, this corresponds to the estimator

\begin{equation}
    \langle p_{\varnothing\to A}^{(q)}\rangle_{\text{MC}} = \frac{1}{N_\text{MC}}
    \sum_\text{\text{MC}}
    \frac{1}{\Lambda} \sum_i
    \delta_{\ket{i_A^{(1)}},\ket{i_A^{(2)}}}\dots\delta_{\ket{i_A^{(q-1)}},\ket{i_A^{(q)}}},
    \label{eq:naive_estimator}
\end{equation}

where the first sum runs over the Markov chain of length $N_\text{MC}$ and the second sum over all
$\Lambda$ slices $i$ of the operator strings in the $q$ replicas. For simplicity, we enforce the
same cutoff $\Lambda$ for all replicas.

This method can be greatly enhanced by the observation that all operator strings in the ensemble
$\{\varnothing\}$ are \emph{independent}. This leads to two possible improvements:
\begin{itemize}
    \item Due to the cyclicity of the trace, each
operator string has a cylinder topology and can thus be independently translated cyclically by any number of states in the
``imaginary time'' direction. Each of the such transformed Monte Carlo configurations has exactly
the same weight.
\item  If the system is invariant under (spatial) symmetry transformations, we can transform the \emph{whole}
    operator string of each replica with \emph{independent} transformations without changing the
    weight of the configuration.
\end{itemize}
These two recipes can be used to greatly improve the quality of the estimation of $S_q^\text{PR}$.
However, a naive application of these ideas is not possible as it is far too expensive to try all
combinations of shifted and transformed operator strings. As discussed below, it is possible to
exploit all symmetries by only one pass on each operator string. 

\subsection{PR entropy of the full system $A=\overline{\varnothing}$}
Let us first start with the full system $A=\overline{\varnothing}$ as the symmetries are simpler to
apply and the resulting improved estimator formulae are clearer.

For each state $\ket{i^{(\alpha)}}$ in the operator string of replica $\alpha$, we calculate the
parent state $p(\ket{i^{(\alpha)}})$ by applying all model symmetries. The application of all
symmetries classes all the basis states $\ket{i}$ into \emph{nonoverlapping} families of states, each
of which is represented by the uniquely defined parent state -- here, we will take the state in the family with the
smallest binary representation. It is important to record the multiplicity (\ie the number of states
belonging to the family) $d(\ket{p})$ of each
state family for the purpose of correct normalization. Note that most state families have the
maximal multiplicity given by the number of symmetries $n_\text{sym}$, with the exception of high
symmetry states for which $d<n_\text{sym}$.

While transversing all the operator strings, we record the histogram $h=\{n(\ket{p},\alpha)\}$ of the number of occurrences
$n(\ket{p},\alpha)$ of states with parent $\ket{p}$ in operator string $\alpha$.

In the next step, for each parent state that has been observed in one of the replicas we have to
count the number of configurations (of symmetry-transformed operator strings) in which we can find
identical states in all $q$ replicas. It is clearly given by\begin{equation}
    \label{eq:nid}
    n_\text{id}(\ket{p}) = d(\ket{p}) \prod_{\alpha=1}^{q}  n(\ket{p},\alpha).
\end{equation}
Note that the multiplicity $d(\ket{p})$ of the parent state $\ket{p}$ accounts for the fact that we
can have any of the $d(\ket{p})$ states in the family of parent $\ket{p}$ as the identical state in
all replicas.

Finally, we have to normalize equation \eqref{eq:nid} by the total number of symmetry equivalent
configurations of all $q$ replicas. As the family of parent $\ket{p}$ consists of
$d(\ket{p})$ states, the correct normalization is $\frac{1}{d(\ket{p})^q}$. Together with
the $\Lambda^q$ possible cyclical shifts of the operator strings, 
this yields the improved estimator in Monte Carlo configuration
$\mathcal{C}$ for the probability of moving
from the independent $\{\varnothing\}$ ensemble to the glued ensemble $\{A\}$ 
\begin{equation}
    \label{eq:improved}
    \langle \langle p_{\varnothing \to \overline\varnothing}^{(q)}\rangle \rangle_{\mathcal{C}} =
    \frac{1}{\Lambda^q} \sum_{\ket{p}\in h}
    d(\ket{p})^{1-q} \prod_{\alpha=1}^{q}  n(\ket{p},\alpha).
\end{equation}
Here, $\Lambda$ is the number of states in the operator string (forced to
be equal in all replicas for simplicity). The sum runs over all parent states recorded in the histogram $h$. In
equation \eqref{eq:improved} it is immediately clear why this method is extremely beneficial for the
observation of small probabilities: the normalization factor can become a very small number, as
typically the number of applied symmetries $\approx N$ for a translationally invariant system with $N$ sites and
$\Lambda\approx10^3\dots10^5$ in our calculations. For large values of $q$, very small numbers can be
obtained by \emph{one} Monte Carlo measurement and the variance of the estimator is greatly reduced.

Because of the tremendous number of possible combinations of symmetry transformed replica operator
strings, it should be noted that the evaluation of Eq.~\eqref{eq:improved} may cause numerical problems.
The products of the numbers $n(\ket{p},\alpha)$ for all $q$ replicas can easily become too large to
be stored as 64-bit integers. One solution for this problem is to use extended precision floating
point numbers. We have found, however, that it is sufficient to perform the products using
double precision floating point numbers and
performing the sum using Kahan's summation algorithm\cite{kahan_pracniques_1965} to avoid precision loss and cancellation effects.

An interesting further improvement of the method stems from the fact that a \emph{single} simulation of
$q_\text{max}$ replicas can be used for the calculations of  PR entropies $S_{A,q}^{\text{PR}}$ (or equivalently the
probabilities $p_{\varnothing\to A}^{(q)}$) with $q$ ranging from $2$ to $q_\text{max}$. Indeed, at
the time of performing a Monte Carlo measurement, we can select any combination (without repetition) of $q$ replicas out
of the $q_\text{max}$ copies and apply Eq.~\eqref{eq:improved} without the need of creating a new
histogram. This can be repeated for all $\binom{q_\text{max}}{q}=\frac{q_\text{max}!}{q!(q_\text{max}-q)!}$ possibilities,
further improving the precision of the estimate:

\begin{equation}
    \label{eq:improved_combinations}
\begin{split}
    \langle \langle p_{\varnothing \to \overline \varnothing}^{(q)}\rangle \rangle_{\mathcal{C}} &=
    \frac{1}{\binom{q_\text{max}}{q} } \frac{1}{\Lambda^q} \times \\ 
    &\sum_{\ket{p}\in h}
    d(\ket{p})^{1-q} \sum_\gamma  \prod_{\alpha=1}^{q}  n(\ket{p},\gamma(\alpha)).
\end{split}
\end{equation}

where the additional sum runs over all combinations $\gamma$ of the $q_\text{max}$
replicas.

\subsection{PR entropy of subsystem $A$}

  If the subsystem $A$ and the full system are not identical, we have to slightly modify  the procedure
described above. The reason for this is the fact that if we cut out the part of subsystem $A$ from
every state in the family corresponding to a parent $\ket{p}$ and for a different family
corresponding to $\ket{p'}$, we will find that the families can now have an overlap if the full
system has more or different symmetries than the subsystem, which is generally the case. In some
cases, where the exploited symmetries of the subsystem are identical with the symmetries of the full
system, the same algorithm as for the full system may be used. This is generally the case if any
symmetry transformation maps the subsystem on itself, \eg in the case of the periodic ladder (see below, Sec.~\ref{sec:ladd}) when solely using translation symmetries along the ladder.

In the general case, accounting for
these overlaps is expensive as all pairs of parent states in the histogram $h$ have to be treated.
We therefore choose to go through all parents in the histogram $h$ and create the family of states
from which we deduce the subsystem states $\ket{i}_A$ by cutting out the corresponding part. In doing so,
we generate a new histogram $h_A$ filled with the cut states $\ket{i}_A$ where we accumulate the corresponding
$n(\ket{p},\alpha)$ from histogram $h$. Note that the histogram $h_A$ may become very large if the
number of symmetries and the number of lattice sites in subsystem $A$ is large. The size of the
histogram $h$ is always smaller or at most equal to $q_\text{max} \Lambda$ (typically, its size is reduced by the number $n_\text{sym}$ of symmetries to $q_\text{max} \Lambda/n_\text{sym}$). 
The maximal size of the histogram $h_A$ is,
however, given by $\max( n_\text{sym} q_\text{max} \Lambda , \mathcal{N}_A)$, where $\mathcal{N}_A$ is
the dimension of the Hilbert space of states on subsystem $A$. Note that the relevant number of symmetries $n_\text{sym}$ here is the number of applied symmetries
of the full system as we discuss here the general case, in which the subsystem has no (or less) symmetries.

The equation for the estimator in the previous section only have to be slightly modified:
\begin{equation}
    \sum_{\ket{p}\in h} d(\ket{p})^{1-q}  \rightarrow \frac{1}{n_\text{sym}}\sum_{\ket{i}_A \in h_A}, \quad \text{and}\quad
    \ket{p}\rightarrow \ket{i}_A,
\end{equation}
where $n(\ket{i}_A,\alpha)$ is then the number of times the subsystem state $\ket{i}_A$ has been
observed in all symmetry realizations of replica $\alpha$. This yields the final estimator
\begin{equation}
    \label{eq:improvedA_combinations}
\begin{split}
    \langle \langle p_{\varnothing \to A}^{(q)}\rangle \rangle_{\mathcal{C}} &=
    \frac{1}{\binom{q_\text{max}}{q} } \frac{1}{(n_\text{sym} \Lambda)^q} \times \\ 
    &\sum_{\ket{i}_A \in h_A}
    \sum_\gamma  \prod_{\alpha=1}^{q}  n(\ket{i}_A,\gamma(\alpha)).
\end{split}
\end{equation}

\subsection{Autocorrelation problem for large values of $q$}

  While the improved estimator performs remarkably well for a wide range of $q$ and compares
perfectly with exact results for small systems (see Sec.~\ref{sec:results}), we have found that it may yield wrong results for
large values of $q$ in some extreme cases. A detailed investigation shows that this
behavior stems from an increasing variance of the improved estimator with $q$ together with an
increasing autocorrelation time which may exceed the simulation time and thus yield systematic
errors\footnote{It should be noted that this problem only occurs in the regime of very large PR
    entropies, far beyond the reach of the simple estimator given in Eq.~\eqref{eq:naive_estimator}.}.

\begin{figure}[h]
    \centering
    \includegraphics{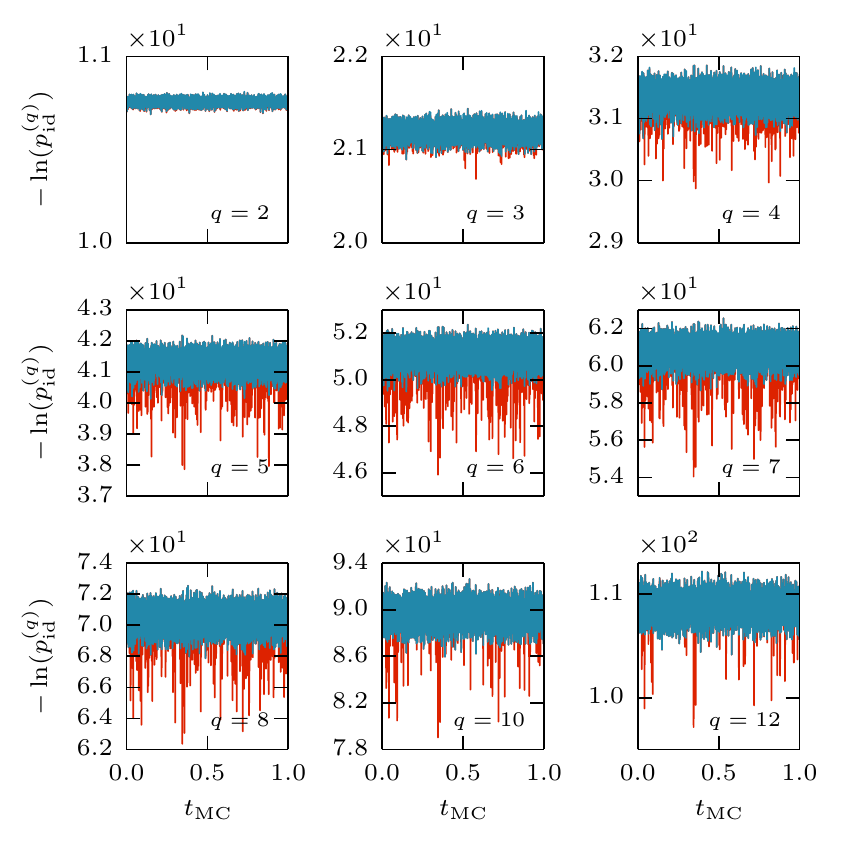}
    \caption{(Color online) Comparison of the time series of (minus log of) the estimator given by Eq.~\eqref{eq:improvedA_combinations} (shown in red) and the estimator excluding the most
        probable states (shown in blue) for different R\'enyi indices $q$ for the $L=16$ Heisenberg
    ladder at $J_\perp=4$. Clearly, the variance of the estimator excluding the most probable states
is very much reduced.}
    \label{fig:improvedestimator}
\end{figure}

The reason for this behavior is identified by studying the time series of the estimator for
  different values of $q$ (see Fig.\ref{fig:improvedestimator}). Clearly, for larger values of
  $q$, the time series shows more and more pronounced ``spikes'' that at the same time become rare
  and large for large $q$. We have found that the most severe spikes are created by the occurrence
  of the most probable states (for antiferromagnetic systems, these are the N\'eel states
  $\ket{N_A}$ and $\ket{N_B}$, see
  Refs.~\onlinecite{luitz_universal_2014,luitz_shannon_2014,luitz_participation_2014}). These states
  occur with the highest probability $p_\text{max}$ in any operator string and have an enhanced
  symmetry (for instance the symmetry family of the N\'eel states only has a $d=2$ multiplicity).
  Consequently, the factor $d(\ket{p})^{1-q}$ in Eq.
  \eqref{eq:improved_combinations} is much
  larger than for less symmetric states (for which it is usually $n_\text{sym}^{1-q}$), creating
  extremely large spikes for large $q$ if the N\'eel states are observed in all $q$ replicas simultaneously.
  The frequency of the spikes becomes however rare with growing $q$, as the simultaneous observation
  probability of the N\'eel states in all $q$ replicas decreases rapidly as $p_\text{max}^q$.

  However, we can remedy this situation by calculating the probability $p_\text{max}$ separately, recording
  the frequency of the N\'eel states across all replicas. We then deliberately exclude the most
  probable states with common parent $\ket{N_A}=p(\ket{N_A})=p(\ket{N_B})$ from the estimator in Eq.~\eqref{eq:improved_combinations} (similarly for Eq.~ \eqref{eq:improvedA_combinations}) and obtain
  \begin{equation}
 \begin{split}
   & \langle \langle p_{\varnothing \to \overline \varnothing}^{(q)}\rangle \rangle_{\mathcal{C}} =
    \frac{1}{\binom{q_\text{max}}{q} } \frac{1}{\Lambda^q} \times \\ 
    &\sum_{\ket{p}\in h,\ket{p}\neq\ket{N_A}}
    \left[    d(\ket{p})^{1-q} \sum_\gamma  \prod_{\alpha=1}^{q}  n(\ket{p},\gamma(\alpha)) \right]
    \\ &+ d(\ket{N_A})
    p_\text{max}^q.
\end{split}
      \label{eq:improvedpmax}
  \end{equation}
  In order to pull out the term $d(\ket{N_A})p_\text{max}^q$ out of the sum, we have used the fact
  that on average the frequency $n(p(\ket{N_A}),\alpha)$ of the
  N\'eel states in replica $\alpha$ of length $\Lambda$ is given by $\Lambda d(p(\ket{N_A}))
  p_\text{max}$.
  
  In Fig.~\ref{fig:improvedestimator}, we compare the time series of (minus log of) the estimator given by
  Eq.~\eqref{eq:improved_combinations} and the improved estimator (excluding the additive term
  $+ d(\ket{N_A}) p_\text{max}^q$) given by Eq.~\eqref{eq:improvedpmax} in a test example discussed deeper later. Clearly, for $q=2$ the two estimators behave nearly
  exactly equally. However, as $q$ grows, the estimator not excluding the most probable states
  (shown in red) has a much larger variance than the estimator that excludes the most probable
  state. In addition to that, the spikes created by the N\'eel state become rare with growing $q$
  and eventually may not be even recorded once in a simulation, leading to incorrect results (especially since $p_\text{max}$ will dominate the PR entropy
  for large values of $q$). 
  
Note that usually $p_\text{max}$ can be calculated with a much higher precision than any other state, leaving us with the possibility to correct efficiently this systematic error. Any other state can be taken out of the estimator as in Eq.~\eqref{eq:improvedpmax} if its
  probability can be measured to a sufficient accuracy directly. This way, a hybrid method capturing
  the histogram of the most probable states with high precision and calculating the correction
  caused by the less probable states by the replica trick can be easily constructed. 

\subsection{Measurement of the replica correlation entropy}

The final element to compute the entanglement entropy is the measurement of the replica correlation entropy $C_{A,q}^{\text{R}}$, for which the probability $p_{A\to \varnothing}$ has
to be estimated efficiently. 

In the method of Ref.~\onlinecite{humeniuk_quantum_2012}, the condition for moving from the glued ensemble to the independent
ensemble is given by $\ket{0_A^{(2)}} = \ket{0_A^{(1)}}$. In this case, the glue can be cut and
rewired thus moving from the glued to the independent ensemble \emph{without changing the
weight} as illustrated in Fig. \ref{fig:glued}.

In our scheme, we can simply simulate the glued ensemble and measure the number of times we observe
$\ket{0_A^{(2)}} = \ket{0_A^{(1)}}$ relative to the total number of QMC steps. Unfortunately, the glued replicas can not be changed by separate symmetry transformations and the
topology of the glued operator string is rigid in imaginary time, \ie it can not be translated. 
Generally, it is therefore not possible to construct an improved estimators for
$C_{A,q}^{\text{R}}$ in the same way as for $S_{A,q}^{\text{PR}}$.

\section{Thermodynamic R\'enyi entropies}
\label{sec:renyi}

  Let us discuss in more detail the special case in which the subsystem $A$ and the full system
$\overline{\varnothing}$ are identical. In that case, the entanglement R\'enyi entropies
$S_q^\text{E}$ reduce to the thermodynamic R\'enyi entropies $S_q(\beta)$ at inverse temperature
$\beta$, which are defined by:
\begin{eqnarray}
    S_q^{\rm th}(\beta) &=& \frac{1}{1-q} \ln \frac{\tr~\E^{-q\beta \hat{H}}}{\left(\tr~\E^{-\beta
        \hat{H}}\right)^q}\nonumber\\
        & =& \frac{q\beta}{1-q} \left[ F(\beta) - F(q \beta) \right].
    \label{eq:thermalSq}
\end{eqnarray}
Here, $F(\beta)=-\frac{1}{\beta} \ln \tr~\E^{-\beta \hat{H}}$ denotes the free energy of the system
governed by the Hamiltonian $\hat{H}$ at inverse temperature $\beta$. Noting that the thermal
density matrix $\rho_{\overline{\varnothing}}$ is nothing else but the reduced density matrix of the
subsystem $A=\overline{\varnothing}$, we can use Eq.~\eqref{eq:EEfromPR} to calculate the
thermodynamic R\'enyi entropy.

  As in the case of the entanglement R\'enyi entropy, we will decompose the thermodynamic R\'enyi
  entropy in a difference of the participation PR entropy and the replica correlation. The discussion of Sec.~\ref{sec:PR} on how to build improved estimators exploiting spatial and imaginary time symmetries carries on for the calculation of the PR
  entropy. 
  
The calculation of the replica correlation for  $A=\overline{\varnothing}$ can be much further improved than in the generic case. Indeed this is the case where the replicas are glued on the full system and the periodicity in $\beta$ (or in the cutoff $\Lambda$ for SSE) is replaced
  by a $q \beta$ periodicity for a larger operator string. Clearly, this situation can be achieved
  by simply performing a standard SSE simulation with a {\it single replica} at inverse temperature
  $q\beta$. The estimator for the replica correlation $C_q^\text{R}$ is then given by the
  probability of cutting the enlarged configurations in $q$ valid SSE replicas at inverse temperatures $\beta$, with the condition that  these $q$ replicas have to be periodic in $\beta$.

It should be observed that there are multiple valid ways of slicing the large configuration in $q$ parts
  and in fact any partition $(\Lambda_1,\dots,\Lambda_q)$ is valid \emph{as long as every cutoff
  $\Lambda_i$ is large enough such as to correctly sample the inverse temperature $\beta$}. The probability of slicing the large replica in $q$ valid parts can be estimated (see Appendix \ref{sec:appendix_thermal} for a detailed derivation) by
  simply measuring the observable
\begin{equation}
    \label{eq:thermal_estimator}
    X_{\overline{\varnothing}\to \varnothing} =
    \delta_{\alpha_{1,\Lambda_1},\dots,\alpha_{q,\Lambda_q}}  \frac{ \Lambda! \prod_{i=1}^{q}
(\Lambda_i-n_i)!}{ q^n (\Lambda-n)!
       \prod_{i=1}^q \Lambda_i!}
\end{equation}
for any Monte Carlo configuration in the replica simulated at inverse temperature $q\beta$. 
Here, the Kronecker delta yields $1$ if the states $\alpha_i$ at the beginning of each replica of
length $\Lambda_i$ are identical, which is precisely the condition for obtaining $q$ valid,
$\Lambda_i$-periodic replicas after performing the cut.
Note that one should of course average over several possible partitions $(\Lambda_1,\dots,\Lambda_q)$
with $\Lambda=\sum_i \Lambda_i$ and all translations of the partitions in imaginary time.

We obtain
\begin{equation}
    C_q^\text{R} = \frac{1}{1-q} \ln(
    \langle{X_{\overline{\varnothing}\to\varnothing}}\rangle_{\text{MC},\{\Lambda_i\}})
    \label{eq:cq_thermal}
\end{equation}
where the average is performed over the Markov chain and different partitions in order to improve
the statistics. The thermodynamic R\'enyi entropy is thus finally given by 
\begin{equation}
    S_q^{\rm th} = S_{q}^{\text{PR}} - C_q^\text{R},
    \label{eq:thermal_Sq}
\end{equation}
where the absence of the subsystem index indicates that the full system is to be considered. We emphasize that both $S_{q}^{\text{PR}}$ and $C_q^\text{R}$ are obtained within a standard SSE independent ensemble simulation.

\section{Results}
\label{sec:results}

In this section we present various results obtained on simple model Hamiltonians, such as Heisenberg chains and ladders, in order to carefully test the method. We compare, when possible, our QMC estimates with exact diagonalization (ED) or DMRG results. The quantities of interest we discuss in the rest are the R\'enyi entanglement (zero temperature) and thermodynamic (finite temperature) entropies. We also compare the efficiency of our new method for calculating entanglement entropies $S_{A,q}^{\text{E}}$ with the method of Ref.~\onlinecite{humeniuk_quantum_2012}. 

The last part of this section deals with a careful analysis of the reconstruction of the entanglement spectrum given entanglement entropies. We finally provide a quantitative analysis of how to probe an ansatz entanglement Hamiltonian in the case of Heisenberg ladders.

\subsection{Heisenberg chain}
\label{sec:chain}
As a first application of the different aspects of the method introduced in Sec.~\ref{sec:RC} we
perform calculations for the well-studied antiferromagnetic Heisenberg chain of $L$ spins $S=\frac{1}{2}$, described by the Hamiltonian
\begin{equation}
    H_{\text{1d}} = J\sum_i \vec{S}_i \cdot \vec{S}_{i+1},
\end{equation}
using periodic boundary conditions $\vec{S}_{L+1}=\vec{S}_1$.

\subsubsection{Entanglement entropies}
\begin{figure}[b]
    \centering
    \includegraphics[width=\columnwidth]{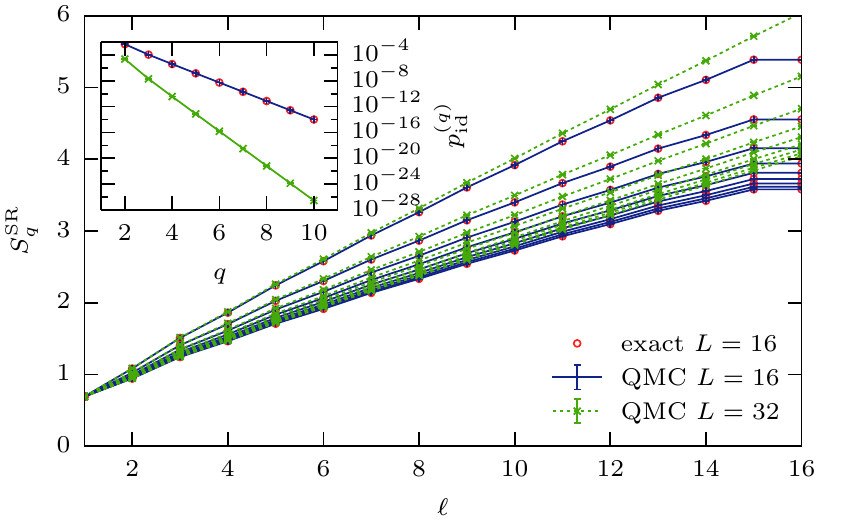}
    \caption{(Color online) QMC results for the $L=16$ and $L=32$ Heisenberg chains. We also show the comparison to the exact solution obtained by
        exact diagonalization for $L=16$. The main panel shows the PR entropy $S_q^{\text{PR}}$ as a function
        of subsystem size $\ell$ for
        $q=2,3\dots10$, where $q=2$ corresponds to the largest entropies and $q=10$ to the smallest
        ones. In the inset, we display the corresponding probability $p_\text{id}^{(q)}$ of finding the
        same state for the full system $\ell=L$ in all $q$ replicas.
        The QMC results stem from $10^6$ Monte Carlo measurements and the average expansion order
        was $\langle n \rangle = 891.36(2)$ for $L=16$ and $\langle n \rangle = 3553.08(5)$ for
        $L=32$. Using a simple average over all imaginary time slices,
        one would be blind to probabilities $p\lesssim10^{-9}$. With our improved method, we not only are
        able to calculate \emph{much} smaller probabilities, but also obtain a very small variance
        of the result.
}
    \label{fig:shannon_xxx}
\end{figure}

We test our implementation on the example of a chain of $L=16$ spins,
which can be easily solved using ED. As a way to show the different elements in our method, we first display in the main panel of Fig.~\ref{fig:shannon_xxx} the participation entropies $S_q^{\text{PR}}$ as a function
of the subsystem size $\ell$ and R\'enyi index $q$. The correspondence between the QMC and exact
result is perfect and the precision is such that the error bars are not even visible in the graph. The inset displays the
probabilities $p_\text{id}^{(q)}$ of finding the same basis state in $q$ replicas for $q$ ranging
from 2 to 10, for the full system $\ell=L$. The exponential decay of $p_\text{id}^{(q)} = \exp[-(q-1) S_q^\text{PR}]$ is perfectly
reproduced by the Monte Carlo result and even the smallest probabilities of the order of $10^{-27}$
are estimated with extremely high accuracy (errorbar of the order of $10^{-29}$) in a calculation
containing $10^6$ Monte Carlo measures. In order to appreciate this result, let us mention that the simple estimator given in Eq.~\eqref{eq:naive_estimator} can hardly see a single event occurring with such low probabilities. The reason for this is that the total denominator $N_{MC} n$ for the present case is given by
$N_{MC}=10^6$ and the expansion order $\langle n \rangle = 3553.08(5)$ (for the case of $L=32$), 
thus it is of the order of $10^9$. Consequently, events of probabilities $p<10^{-9}$ will typically never be seen in a
Markov chain of length $N_{MC}\approx 10^6$.

\begin{figure}[t]
    \centering
    \includegraphics{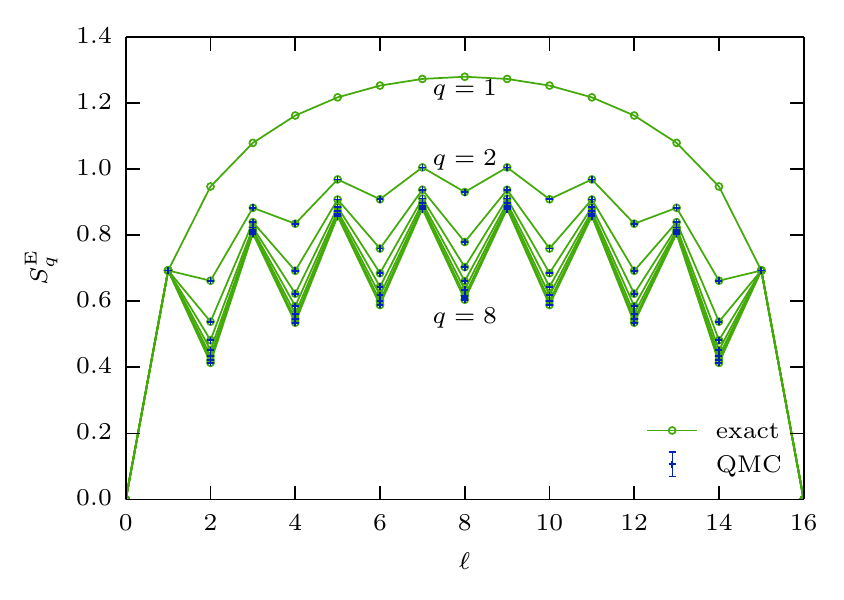}
    \caption{(Color online) QMC results for the entanglement R\'enyi entropies $S_q^\text{E}$ for a bipartition of the
    $L=16$ Heisenberg chain as a function of subsystem size $\ell$ compared to exact
    diagonalization. The Monte Carlo results have been obtained from Eq.~\eqref{eq:EEfromPR}
and have been symmetrized around $\ell=L/2$. The $q=1$ result is not accessible from a Monte Carlo
calculation and is included for illustration purposes.}
    \label{fig:xxxL16EE}
\end{figure}

Let us now combine the result in Fig.~\ref{fig:shannon_xxx} with the result for the calculation of
the replica correlation entropy to obtain the entanglement R\'enyi entropy according to equation
\eqref{eq:EEfromPR}. Fig.~\ref{fig:xxxL16EE} shows the comparison of this result with the one
obtained by ED. The correspondence is again perfect and shows that the method works very well: for instance, the even-odd oscillations for $q\ge 2$ are perfectly reproduced~\cite{Calabrese_oscillations_10}.
Note that the errorbars of the two Monte Carlo results yield the final error of
$S_q^\text{E}$ by
\begin{equation}
    \sigma_\text{E} =  \sqrt{ \sigma_{\text{PR}}^2  + \sigma_{\text{R}}^2}.
\end{equation}
We usually find that the error bar of the replica correlation entropy $\sigma_{\text{R}}$ is larger than
the error of the PR entropy $\sigma_{\text{PR}}$ due to the lack of an improved estimator of
$C_q^{\text{R}}$ and therefore dominates the total error.

\subsubsection{Thermodynamic R\'enyi entropies}

\begin{figure}[t]
    \centering
    \includegraphics{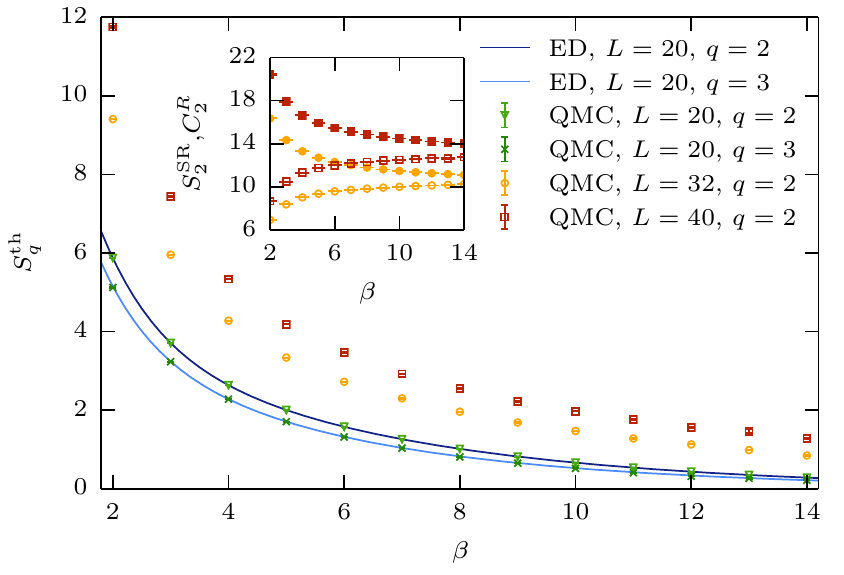}
    \caption{(Color online) Thermodynamic R\'enyi entropies obtained from Monte Carlo simulations. The match with
    the exact diagonalization result is perfect. The inset shows the dependence of the participation entropy
    (filled symbols) and the replica correlation (open symbols) on the inverse temperature $\beta$
for $L=32$ (yellow circles) and $L=40$ (red squares). }
    \label{fig:th_renyi}
\end{figure}

We here now consider the full-system made of the chain to obtain the thermodynamic R\'enyi entropy. We have performed calculations with $q=2,3$ replicas for the
periodic Heisenberg chain at different finite temperature for different system sizes and calculated the participation 
entropies $S_q^{\text{PR}}(\beta)$. A second set of simulations at inverse temperatures $q\beta$ has then
been carried out in order to obtain the replica correlation $C_q^\text{R}$ using Eq.~\eqref{eq:cq_thermal}. We extracted the thermodynamic R\'enyi entropies $S_2^{\rm th}(\beta)$ and $S_3^{\rm th}(\beta)$ from the
result and compare to ED in Figure \ref{fig:th_renyi}. Clearly the QMC
result for $L=20$ matches the exact values perfectly. Furthermore, one can access much larger chain sizes as compared to ED techniques, limited to $L\sim 20$ for the full diagonalization required to access finite temperature behaviour.

We also show the temperature dependence of the individual terms (see inset of Fig.~\ref{fig:th_renyi}) from which the thermodynamic R\'enyi
entropy is obtained. While the participation entropy decreases with inverse temperature (it assumes its maximal
value of $L\ln 2 $ at $\beta=0$), the replica correlation increases to eventually match the
value of the PR entropy at zero temperature.

One can also test with $S=1/2$ chains the conformal field theory prediction for $S^{\rm th}_{q}(\beta)$. In the regime $1\ll u\beta\ll L$ (and ignoring logarithmic corrections due to marginal operators~\cite{Cardy86}), the free energy obeys the following scaling~\cite{Affleck86}
\begin{equation}
F(T)=E_0-L\frac{\pi c}{6u}T^2,
\end{equation}
where $E_0$ is the ground-state energy, $c$ the central charge, and $u$ the velocity of excitations. Using Eq.~\eqref{eq:thermalSq}, one arrives for the low temperature scaling to:
\begin{equation}
S_q^{\rm th}(T)=\frac{\pi c}{6 u}\left(1+\frac{1}{q}\right)LT.
\label{eq:SqT}
\end{equation}

\begin{figure}[b]
    \centering
    \includegraphics{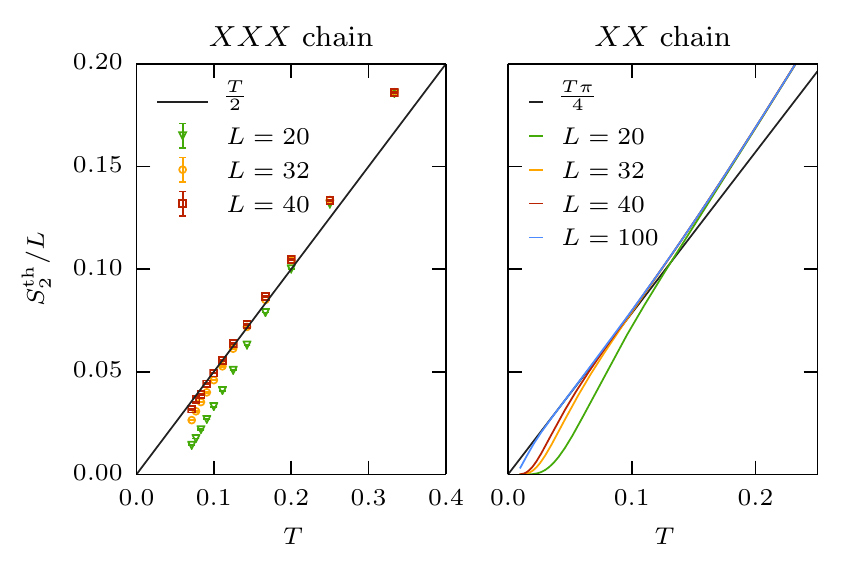}
    \caption{(Color online) Thermodynamic R\'enyi entropies for the XXX (QMC left) and XX (ED right) chains compared to
    the low temperatures CFT prediction Eq.~\eqref{eq:SqT}.}
    \label{fig:cftcomp}
\end{figure}
 
This behavior is checked with $q=2$ for XXX chains of various lengths $L=20,~32,~40$ in Fig.~\ref{fig:cftcomp} (left), where the low temperature linear form is well reproduced using $c=1$ and $u=\pi/2$. 
Finite size convergence effects are due to the finite length gap $G(L)\simeq u/L$ such that the asymptotic low T behavior Eq.~\eqref{eq:SqT} is expected to be valid for $u\gg T\gg G(L)$. Below this gap, $S_q^{\rm th}$ displays an activated shape, controlled by $G(L)$. We checked this finite-size effect using ED at the free-fermion point (XX chain) with open boundary conditions\footnote{The use of open chains simplifies the problem of computing thermal averages in the free fermion representation of the XX spin-$\frac{1}{2}$ chain.} where the asymptotic linear scaling is perfectly well reproduced for large enough sizes $L$, as displayed in Fig.~\ref{fig:cftcomp} (right).

\subsection{Heisenberg ladders}
\label{sec:ladd}
\begin{figure}[h]
    \centering
    \includegraphics[width=3cm]{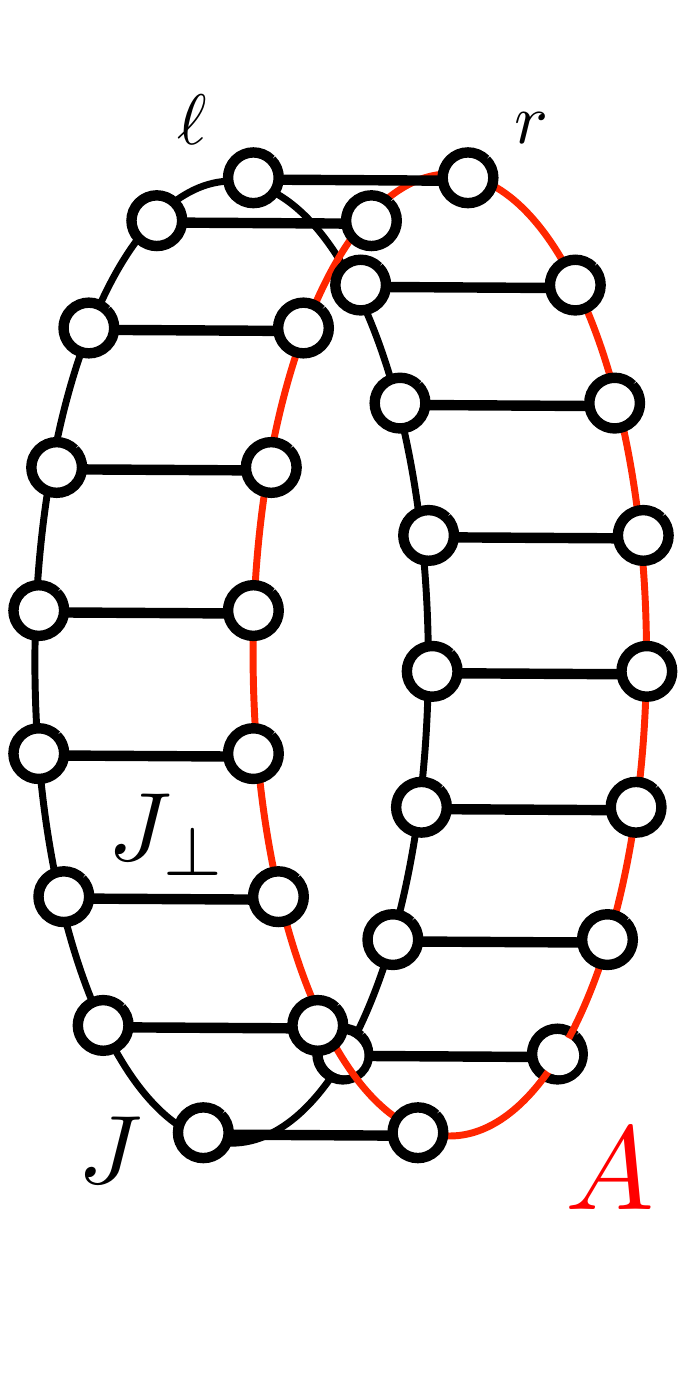}
    \caption{(Color online) Schematic picture for the spin ladder system Eq.~\eqref{eq:ladderH}. Entanglement is studied between subsystem $A$ (red) and the rest.}
    \label{fig:setup}
\end{figure}

  Let us now consider Heisenberg ladders consisting of two
  neighboring one dimensional periodic Heisenberg chains (the ``legs'') with an additional ``rung''
  coupling between the chains:
  \begin{equation}
      H_\text{ladder} = J \sum_{i,\alpha} \vec{S}_{i,\alpha} \cdot \vec{S}_{i+1,\alpha} +
      J_\perp \sum_i \vec{S}_{i,\ell} \cdot \vec{S}_{i,r}.
      \label{eq:ladderH}
  \end{equation}
    where $\vec{S}_{i,\alpha}$ is the spin operator on site $i$ of chain $\alpha=l,r$, corresponding
    to the left and right leg respectively (see Fig.~\ref{fig:setup}). We use periodic boundary conditions along the legs. 
    
For the calculation of entanglement properties, we consider the cut where $A$ is one leg of the ladder and perform calculations in the strongly gapped rung-singlet regime $J_\perp \gg J$, where entanglement entropies are known to be quite large from ED studies~\cite{poilblanc_entanglement_2010,lauchli_entanglement_2012}. The motivation for this regime is to test our method in a difficult, large-entanglement, regime. Such a cut has also been used in several other works on ladder systems \cite{poilblanc_entanglement_2010,Peschel11,cirac_entanglement_2012,lauchli_entanglement_2012,schliemann_entanglement_2012,lundgren_entanglement_2012,Chen13,lundgren_entanglement_2013}.

\subsubsection{Entanglement entropies}

\begin{figure}[t]
      \centering
      \includegraphics[width=\columnwidth]{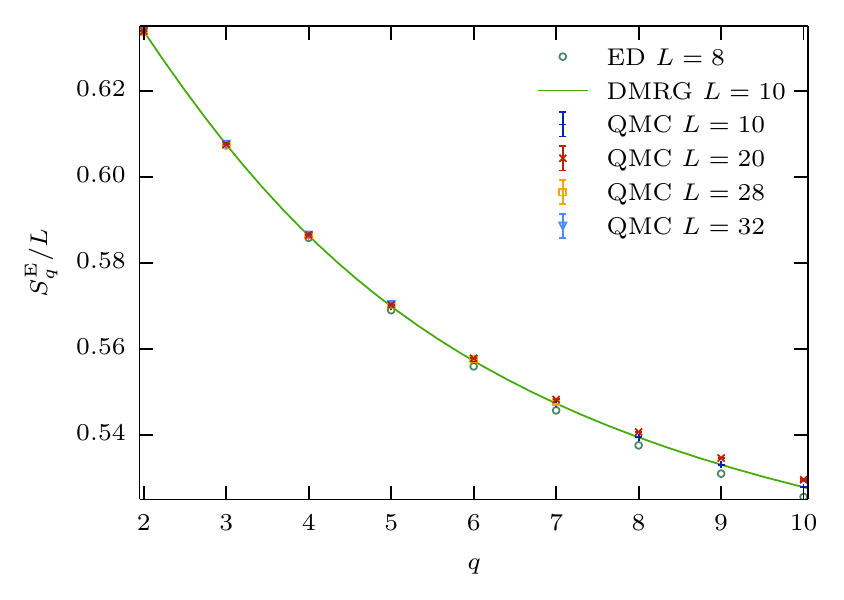}
      \caption{(Color online) Entanglement R\'enyi entropies $S_q^\text{E}$ for different R\'enyi indices $q$ for
      the ladder with $J_\perp=4$. For
  the ladder consisting of $L=10$ rungs, we also add the numerically exact DMRG result for
  comparison. Note that for this particularly strongly entangled system, DMRG can in fact access
  systems up to $L\approx 16$ which corresponds to the limit of
  ED\cite{poilblanc_entanglement_2010}. Our QMC calculation can go further and begins to show
  problems because of too large autocorrelation times around $L=32$ for $q>5$.}
      \label{fig:ladderEE}
  \end{figure}

Fig. \ref{fig:ladderEE} displays our QMC result for various
    values of $q$, system sizes ranging from $L=10$ to $L=32$ and $J_\perp=4J$. For comparison, we also display
    the numerically exact DMRG result for $L=10$. We are still able to perform the calculation for
    $q$ up to as large as $q_{\rm max}=10$ for $L=20$ and begin to see limitations at $q=7$ for $L=28$ as the
    errorbar becomes larger. Clearly, the situation becomes worse for $L=32$, while the result for
    smaller values of $q$ remains extremely good. For comparison, ED (due to the Hilbert space size) or DMRG (due to the large entanglement in this regime) cannot reach systems larger than $L \approx 16$.
    
    Interestingly, the finite size effects on $S_{A,q}^\text{E}/L$ strongly depend on the R\'enyi index $q$.
    For $q=2$, no difference between the result for $L=8$ and the one for $L=32$ is visible,
    however, for $q\ge 7$, $S_{A,q}^\text{E}/L$ displays a sizeable finite length $L$ dependence. This can be easily understood if one realizes that the R\'enyi index $q$ plays the role of an inverse temperature in the entanglement spectrum. This behavior points to a stronger finite size dependence of the lowest lying level
  of the entanglement spectrum (\ie the groundstate energy of the entanglement Hamiltonian --- see discussion later) than for
  high temperature quantities, which are averaged over the whole spectrum.

 \subsubsection{Comparison with the mixed ensemble method}

  In order to get an estimate of the efficiency of the method discussed in this article, we
performed calculations for the $L=20$, $J_\perp=4$ Heisenberg ladder, where subsystem $A$
corresponds to one leg of the ladder (Fig.~\ref{fig:setup}), and compare to results obtained using the method of Humeniuk and Roscilde~\cite{humeniuk_quantum_2012} where for every $q$, we optimized the subsystem increment used for the ratio trick~\cite{hastings_measuring_2010}.

Using the same amount of CPU time ($80\%$ for $C_q^\text{R}$ and $20\%$ for $S_q^\text{PR}$ for our method), we compare
the values of the errorbars between the two different methods. The results are shown in Table
\ref{tab:comp}.
Clearly, for small values of $q$, the
error bars obtained from our method are reduced by one order of magnitude. For
very large values of $q$, the situation changes and the ratio trick provides a better accuracy, leading to an
errorbar that is roughly $2.5$ times smaller. This comes from the fact that
computing the replica correlation $C_q^\text{R}$ part scales exponentially
with $q$, and, as it was explained previously, no improved estimator is
available. For practical purposes, it means that we are limited to values of
$q$ such that $p_{A\to \varnothing} \gtrsim 10^{-7}$ (corresponding to $q=10$ in
this particular case) and the associated error will dominate the total error. 
On the other hand, in the mixed ensemble calculation the ratio trick offers
some flexibility in a certain range of $q$. For $q=2$,
the best error bar is generally obtained for an increment larger that one,
whereas for larger $q$ it becomes quickly much more efficient to set it to
unity. Thus, it becomes computationally more interesting to make several simulations
for which the probabilities are larger, yielding an exponential gain, while the cpu time required by the increasing number
of simulations in order to maintain a constant error bar increases as $l^2$ ($l$ the number of
increments). This explains why the mixed ensemble method~\cite{humeniuk_quantum_2012} becomes
more efficient when $q$ grows.

\begin{table}
    \centering
    \begin{ruledtabular}
    \begin{tabular}{c|cc}
        $q$ & $S_q^E(\text{ratio trick})$ & $S_q^E(\text{from PR})$ \\
        \hline
        $2$  &  $12.67691 \pm 0.00031$ & $12.676998 \pm 0.000026$ \\
        $3$  &  $12.15135 \pm 0.00040$ & $12.151270 \pm 0.000040$ \\
        $6$  &  $11.15622 \pm 0.00076$ & $11.156525 \pm 0.000394$ \\
        $10$ &  $10.58810 \pm 0.00201$ & $10.590639 \pm 0.005246$ \\
    \end{tabular}
    \end{ruledtabular}
    \caption{Entanglement entropy of the $L=20$, $J_\perp=4$ Heisenberg ladder as calculated using
        the method described in Ref. \onlinecite{humeniuk_quantum_2012} (left column) and using the
    method presented in this article (right column). For both calculations we used the same total amount
of CPU time on the same computer. Note that for $q=2$ and $q=3$, our proposed method reduces the
errorbar of the result by roughly a factor of $10$, while for $q=10$, the ratio trick becomes more
efficient. }
    \label{tab:comp}
\end{table}

It should be noted that this comparison is rather rough, as all three implementations have slightly
different optimization goals. Additionally, we did not optimize the CPU time ratio between the
calculation of $C_q^\text{R}$ and $S_q^{\text{PR}}$, which can certainly lead to some improvement.
For the calculation of $S_q^{\text{PR}}$ we used $10$ replicas and obtained the result for
$q=2,3,\dots,10$ in one single simulation, while in the other two simulations, every $q$ has to be
done separately. Therefore the comparison gives a slight advantage to the method of Humeniuk and
Roscilde as the $S_q^{\text{PR}}$ calculation provides more information (on $q=4,5,7,8$ and $9$)
than needed.

  Let us finally mention that the biggest advantage of the method proposed in this article is found
in situations of very large entropies, such as the example of the ladders presented here. For the
case of weak entanglement entropies such as the one dimensional Heisenberg chain, we obtained
roughly the same errorbars in both methods for $q=2$, indicating that the mixed ensemble method performs very
efficiently here.

\subsection{Entanglement spectrum reconstruction}
 
   Having access to the entanglement entropies for various values of the R\'enyi index, we are able
   to explore a recently proposed method for the reconstruction of the entanglement spectrum from
   R\'enyi entanglement entropies measured in QMC\cite{song_bipartite_2012,chung_entanglement_2013}.
   The method relies on the Newton-Girard identities, linking the coefficients of a polynomial to the
   power sums of its roots. This means that a polynomial with roots at the $\lambda_i$ corresponding
   to the entanglement spectrum can be constructed from the knowledge of R\'enyi entanglement entropies ($\lambda_i$ are the eigenvalues of the reduced density matrix).
 
   \begin{figure}[h]
       \centering
       \includegraphics{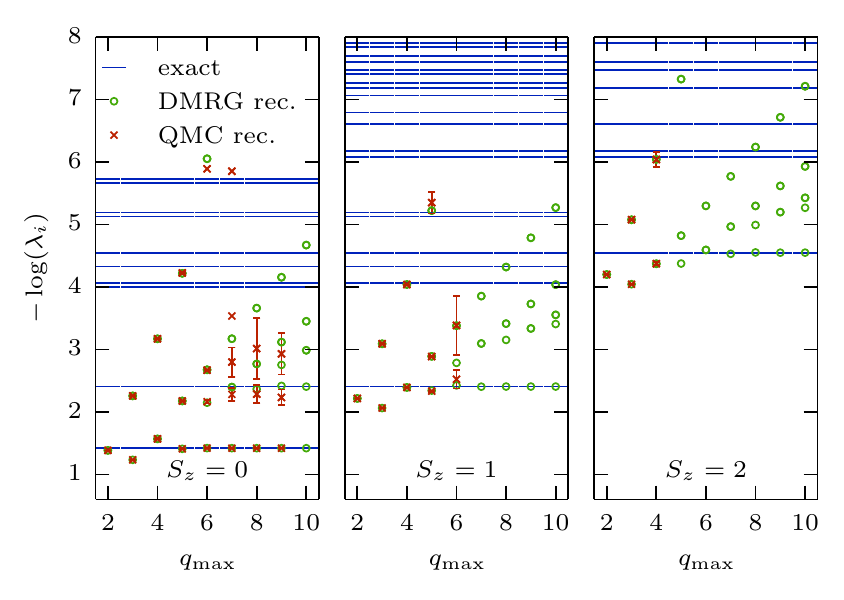}
       \caption{(Color online) Reconstructed entanglement spectrum for the $L=10$ Heisenberg ladder at $J_\perp=1$ as a
           function of $q_\text{max}$. We have separated the spectrum in symmetry sectors of $S_z$
           similar to Ref. \onlinecite{chung_entanglement_2013} and display the exact spectrum obtained
       from a DMRG calculation for reference. In addition to the reconstructed spectrum from our QMC
   data, we also reconstruct the entanglement spectrum from the exact DMRG entanglement entropies in
   each sector in order to study the role of the statistical errors. We have performed the
   reconstruction for different cutoffs $q_\text{max}$ corresponding to the maximal R\'enyi index
   involved in the reconstruction. Errorbars stem from a bootstrap analysis of QMC data. }
       \label{fig:ladderspec}
   \end{figure}
   \begin{figure}[h]
       \centering
       \includegraphics{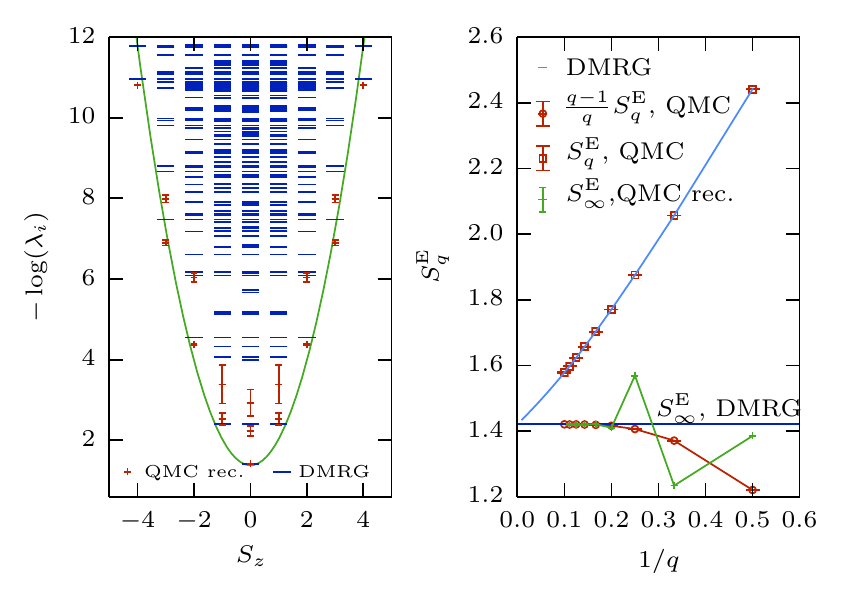}
       \caption{Left: Entanglement spectrum for the $J_\perp=1$ Heisenberg ladder of length $L=10$ obtained
       from DMRG and resolved by symmetry sectors of $S_z$. The parabolic envelope is a guide to the
   eye (see text). Right: R\'enyi entanglement entropy as a function of $1/q$ as obtained from our QMC
   calculation compared to the DMRG result. The horizontal line corresponds to $S_\infty^\text{E}$
   obtained from DMRG. We also show the lowest energy level of the reconstructed entanglement
   spectrum from the QMC data as a function of $1/q_\text{max}$ and the function
   $\frac{q-1}{q}S_q^\text{E}$ from the QMC result. Remarkably, both curves tend to
   $S_\infty^\text{E}$ equally fast, however obtaining $S_\infty^\text{E}$ from the large $q$ limit
   of $\frac{q-1}{q}S_q^\text{E}$ may be more reliable.}
       \label{fig:specsz}
   \end{figure}

   However, in a practical QMC calculation, the R\'enyi entanglement entropies are neither known for
   arbitrarily many values of $q$ nor to unlimited precision. Therefore, the polynomial has to be
   truncated and its order is limited to the maximal R\'enyi index $q_\text{max}$. For the one-dimensional extended
   Bose-Hubbard model, Chung \etal\cite{chung_entanglement_2013} obtained interesting results for
   the low lying entanglement spectrum using $q_\text{max}=4$. Here, we perform a similar
   calculation for the entanglement spectrum of a $L=10$ Heisenberg ladder (see Fig.~\ref{fig:setup}, using the same bipartition as previously) for $J_\perp=J$ and focus on the role of systematic and statistical errors. As in Ref.~\onlinecite{chung_entanglement_2013}, we split the reduced density matrix in its symmetry sectors
   and perform the calculation in each sector (see Appendix \ref{sec:szsector}).

   We have calculated the R\'enyi entanglement entropies up to $q_\text{max}=9$ with the QMC method
   described above for the sectors $S_z=0$, $1$ and $2$ ($q_\text{max}$ decreases with $S_z$ due to
   too large entropies) and reconstructed the entanglement spectrum, systematically varying
   $q_\text{max}$ in order to demonstrate the rate of convergence towards the exact entanglement
   spectrum obtained from DMRG. In Fig. \ref{fig:ladderspec}, we also show the reconstructed
   spectrum from the exact DMRG entanglement entropies. The deviation of the reconstructed DMRG
   spectrum from the exact spectrum gives an impression of the systematic error due to the
   truncation of the polynomial, while the reconstructed QMC spectrum carries additional errors due
   to the statistical uncertainty of the entanglement entropies.

   The convergence of the lowest level (largest eigenvalue $\lambda_0$ of $\rho_A$) in the
   sector $S_z=0$ is very good and the QMC result is trustworthy. This result corresponds to the
   single copy entanglement entropy $S_{A,\infty}^\text{E}=-\ln \lambda_0$ for which no direct QMC estimate is available. 
   The lowest level in the
   sector $S_z=1$ also seems to be converged (within errorbars), however, judging from the QMC data
   only, it is not possible to decide whether the result is trustworthy or not. All other levels can
   not be trusted due to the statistical uncertainty. Therefore, one should bear in mind that a
   careful convergence analysis with $q_\text{max}$ has to be carried out and that in general only
   the lowest part of the entanglement spectrum can be extracted from QMC data bearing statistical
   uncertainties. The reconstructed spectrum from the exact DMRG entanglement entropies shows in
   particular, how slowly higher levels of the entanglement spectrum converge with $q_\text{max}$.
   This is the reason why it is beneficial to split the calculation in symmetry sectors.

A global view on the full entanglement spectrum, resolved in spin sectors, is provided in Fig.~\ref{fig:specsz} (left) where the exact entanglement levels from DMRG are compared to the reconstruction from QMC data. A parabolic enveloppe is shown, as expected from the low energy spectrum of XXZ chains~\cite{alcaraz_conformal_1987}. The right panel of Fig.~\ref{fig:specsz} shows that besides the reconstruction method, it is also possible to accept the single copy entanglement $S_{A,\infty}^\text{E}$ using an extrapolation in $q/(q-1)$ of the R\'enyi entanglement entropies $S_{A,q}^\text{E}$.

   \subsection{Entanglement Hamiltonian}

  Given the difficulty of the extraction of the entanglement spectrum from QMC data, different
  methods of the verification of effective entanglement Hamiltonians and the extraction of the
  inverse entanglement temperature $\beta_\text{eff}$ should be explored. We have proposed the usage of
  the participation spectrum for this purpose in a previous work\cite{luitz_participation_2014},
  which can only provide partial proof that the effective
  model is correct. Here, we propose a different and in fact complementary method that relies on the
  comparison of the R\'enyi entanglement entropy and the finite temperature thermodynamic R\'enyi entropy of a putative entanglement Hamiltonian.

  \begin{figure}[h]
      \centering
      \includegraphics{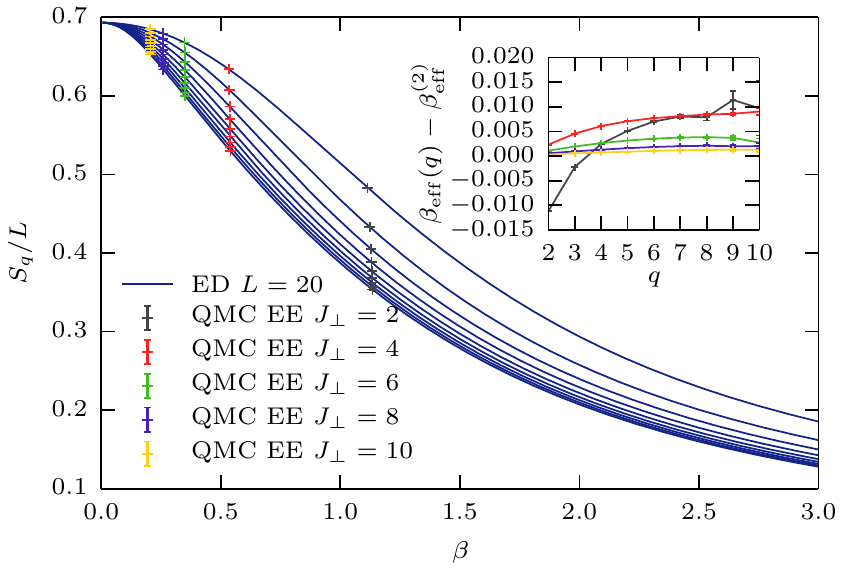}
      \caption{(Color online) Comparison of the thermodynamic R\'enyi entropy of the XXX chain for $q=2$ (top curve) to $q=10$ (bottom
      curve) and the entanglement
      R\'enyi entropy of the Heisenberg ladder using several values of $J_\perp/J$ (coloured symbols). For every value of $q$, we calculate the inverse temperature $\beta$ at which the
      entanglement entropy matches the thermodynamic entropy of the XXX chain. The inset displays
      the deviation of the corresponding effective inverse temperatures $\beta_\text{eff}$ from the
      second order result $\beta_\text{eff}^{(2)}=\frac{2}{J_\perp} + \frac{1}{2 J_\perp^2}$
      (\cf Ref. \onlinecite{schliemann_entanglement_2012}) as a function of $q$ (see main text).
       }
      \label{fig:ladder_thermo}
  \end{figure}

The reduced density matrix $\rho_A$ can be expressed as a thermal mixed state of
    an effective entanglement Hamiltonian $\hat H_\text{E}$ at inverse temperature $\beta_{\text{eff}}$
    by
    \begin{equation}
        \rho_A=\frac{1}{Z} {\E^{-\beta_\text{eff} \hat H_\text{E}}},
        \label{eq:entanglementH}
    \end{equation}
    with $Z=\tr\,\E^{-\beta_\text{eff}
        \hat H_\text{E}}$. Therefore, 
    the thermodynamic entropies $S_q^{\rm th}$ of the effective model $\hat H_\text{E}$ at inverse temperature
    $\beta_\text{eff}$ have to be equal to the entanglement entropies $S_{A,q}^\text{E}$
    \emph{for all $q$}, which is obvious from the above definition Eq.~\eqref{eq:entanglementH} of $\hat H_\text{E}$.

    In the case of the strongly entangled Heisenberg ladder ($J_\perp>J$), it was shown within first
    order perturbation theory \cite{Peschel11,lauchli_entanglement_2012,schliemann_entanglement_2012,Chen13} that the
    entanglement Hamiltonian is given by the simple Heisenberg chain at an effective inverse
    temperature of $2/J_\perp$. The second order correction for the effective temperature (\cf
    Eq. (21) in Ref. \onlinecite{schliemann_entanglement_2012}, ignoring next nearest neighbor interactions) yields
    \begin{equation}
        \beta_\text{eff}^{(2)} = \frac{2}{J_\perp}+\frac{1}{2 J_\perp^2}.
        \label{eq:betaeff_secondorder}
    \end{equation}
  
    In Figure \ref{fig:ladder_thermo}, we explore the regime of validity of the first-order
    entanglement Hamiltonian. We display $S_q^{\rm th}(\beta)$
    of the $L=20$ Heisenberg
    chain for different R\'enyi indices $q$ as a function of inverse temperature $\beta$. Then, we
    calculate the entanglement entropy of the $L=20$ Heisenberg ladder for different values of
    $J_\perp=2,~4,~6,~8,~10$ with our QMC method and extract the effective inverse temperature for which the two quantities match. If
    the entanglement Hamiltonian is correct, the resulting effective inverse temperature has to be
    independent of $q$. This is clearly the case if $J_\perp/J$ becomes large, as visible both in main panel and inset of Fig.~\ref{fig:ladder_thermo}, where the deviation from Eq.~\ref{eq:betaeff_secondorder} is getting smaller and flatter (as a function of $q$) when $J_\perp$ increases.

\section{Conclusion}
\label{sec:conc}

  We have shown that the calculation of entanglement R\'enyi entropies may be split into two
independent Monte Carlo simulations, one of which boils down to a standard calculation of the participation R\'enyi entropy $S_q^{\text{PR}}$ obtained from a simulation of $q$ independent replicas,
while the other part is a ``replica correlation'' entropy $C_q^{\text{R}}$ obtained in a simulation
of $q$ replicas glued together on subsystem $A$. As the PR entropy is a basis dependent quantity,
$C_q^{\text{R}}$ has to be basis dependent, too. Both quantities have to be calculated in the same
basis to obtain the correct entanglement entropy. 

  In a second step, we have developed an improved estimator for the PR entropy, exploiting the fact
that the independent replicas can be transformed independently under imaginary time and space
symmetry transformations leaving the weight of the Monte Carlo configuration invariant. The
number of configurations that are averaged over is therefore multiplied by a number growing
exponentially with the number of replicas $q$ and counteracts the exponential decay of the probability
of finding identical states in all replicas with $q$. This improved estimator allows us to
measure extremely low probabilities, crucial for the calculations of $S_q^E$ for large values of the
R\'enyi index $q$.

  Given that the two terms $S_q^{\text{PR}}$ and $C_q^{\text{R}}$ exhibit a volume law, the realm
of applicability of the method is limited to cases, where $C_q^{\text{R}}$ is not too large. This
is precisely the case in situations where the circumference of subsystem $A$ is identical to its
volume such as for the example of the ladders studied in this article. Here, the largest
contribution to the entanglement entropy stems from the PR entropy, which can be calculated with
very good precision due to the improved estimator.

  For situations where the volume of the subsystem becomes large, such as the half system of a two
dimensional lattice, it is possible to combine the two methods to calculate the entanglement entropy. This would result for example to perform the first increment ({\it{i.e.}} a line-shaped subsystem) with the method presented here with very good precision and start from there
exploiting a ratio trick~\cite{hastings_measuring_2010,deForcrand_2001} using the method introduced in Ref.~\onlinecite{humeniuk_quantum_2012}. This
way, the largest growth of entanglement entropy is dealt with by the improved estimator and the
addition of further lattice sites to the subsystem does not increase the entanglement dramatically,
therefore the ratio trick method is supposed to work very well without accumulating larger errors.

  Let us finally note that the lessons from the improved estimator can certainly be implemented in
the method introduced by Humeniuk and Roscilde\cite{humeniuk_quantum_2012}. If the QMC configuration
is in the independent ensemble, one can check if any of the symmetry transformed replica
configurations introduced here matches the gluing condition. If so, one has to actually perform the
transformation of the whole operator string: in practice, this may turn computationally expensive and one would need to check in which situations the improvement in statistics will be worth the additional computational extra-cost.

\section{Acknowledgements}

  We wish to thank S. Capponi for discussions and for providing test data, D. Poilblanc, S. Pujari
  and S. Wessel for useful related discussions and I. McCulloch for providing access to his
  code~\footnote{See \url{http://physics.uq.edu.au/people/ianmcc/mptoolkit/}} used to perform the test DMRG calculations. Our QMC codes are partly based on the ALPS libraries~\cite{ALPS13,ALPS2}. This work was performed using HPC resources from GENCI (grant x2014050225) and CALMIP (grant 2014-P0677) and is supported by the French ANR program ANR-11-IS04-005-01. We also acknowledge the technical assistance by the IDRIS supercomputer center.

\appendix
\section{Improved estimator of thermal R\'enyi entropies}
\label{sec:appendix_thermal}
In order to be explicit, let us start from Eq.~\eqref{eq:thermalSq} and concentrate on the
ratio of partition functions that have to be calculated. Here, we will use equation (263) from Ref.
\onlinecite{sandvik_computational_2010} for the stochastic series expansion of the partition
function in terms of an operator string $S_\Lambda$ composed of bond Hamiltonians $H_{a(p),b(p)}$ linking
the lattice sites $a(p)$ and $b(p)$.

\begin{widetext}
\begin{equation}
    \frac{Z(q\beta)}{Z(\beta)^q} =  \frac{ \sum_\alpha \sum_{S_\Lambda} (-1)^{n_2} \frac{ (q\beta)^n
    (\Lambda-n)!}{ \Lambda! } \bra{\alpha} \prod_{p=0}^{\Lambda-1} H_{a(p),b(p)} \ket{\alpha}}
    { \sum_{\alpha_1\dots \alpha_q} \sum_{S_\Lambda^1\dots S_\Lambda^q} (-1)^{n_2} \frac{ \beta^n
        \prod_{i=1}^{q} (\Lambda_i-n_i)!}{ \prod_{i=1}^q \Lambda_i!} \prod_{i=1}^{q} \bra{\alpha_i}
        \prod_{p=0}^{\Lambda_i-1} H_{a_i(p),b_i(p)} \ket{\alpha_i}}.
\end{equation}
\end{widetext}

Let us now introduce the observable to compute the PR entropy $S_q^{\text{PR}}$ in the ensemble of
$q$ independent replicas. It is given by the Kronecker delta $\delta_{\alpha_1,\dots,\alpha_q}$.

\begin{widetext}
\begin{equation}
\begin{split}
    \frac{Z(q\beta)}{Z(\beta)^q} &=  \frac{ \sum_{\alpha} \sum_{S_\Lambda^1\dots S_\Lambda^q}
(-1)^{n_2} \frac{ \beta^n
        \prod_{i=1}^{q} (\Lambda_i-n_i)!}{ \prod_{i=1}^q \Lambda_i!} \prod_{i=1}^{q} \bra{\alpha}
        \prod_{p=0}^{\Lambda_i-1} H_{a_i(p),b_i(p)} \ket{\alpha}}
 { \sum_{\alpha_1\dots \alpha_q} \sum_{S_\Lambda^1\dots S_\Lambda^q} (-1)^{n_2} \frac{ \beta^n
        \prod_{i=1}^{q} (\Lambda_i-n_i)!}{ \prod_{i=1}^q \Lambda_i!} \prod_{i=1}^{q} \bra{\alpha_i}
        \prod_{p=0}^{\Lambda_i-1} H_{a_i(p),b_i(p)} \ket{\alpha_i}} \times \\
& \times\frac{ \sum_\alpha \sum_{S_\Lambda} (-1)^{n_2} \frac{ (q\beta)^n
    (\Lambda-n)!}{ \Lambda! } \bra{\alpha} \prod_{p=0}^{\Lambda-1} H_{a(p),b(p)} \ket{\alpha}} 
    { \sum_{\alpha} \sum_{S_\Lambda^1\dots S_\Lambda^q}(-1)^{n_2} \frac{ \beta^n
        \prod_{i=1}^{q} (\Lambda_i-n_i)!}{ \prod_{i=1}^q \Lambda_i!} \prod_{i=1}^{q} \bra{\alpha}
        \prod_{p=0}^{\Lambda_i-1} H_{a_i(p),b_i(p)} \ket{\alpha}} =\\ 
        & \E^{(1-q) S_q^{\text{PR}}} \times
\frac{ \sum_\alpha \sum_{S_\Lambda} (-1)^{n_2} \frac{ (q\beta)^n
    (\Lambda-n)!}{ \Lambda! } \bra{\alpha} \prod_{p=0}^{\Lambda-1} H_{a(p),b(p)} \ket{\alpha}} 
    { \sum_{\alpha} \sum_{S_\Lambda(\Lambda_1,\dots,\Lambda_q)}  \delta_{\alpha_{1,\Lambda_1},\dots,\alpha_{q,\Lambda_q}}  \frac{ \Lambda! \prod_{i=1}^{q} (\Lambda_i-n_i)!}{ q^n (\Lambda-n)! \prod_{i=1}^q \Lambda_i!} 
(-1)^{n_2} 
        \frac{ (q \beta)^n (\Lambda-n)!}
        {\Lambda!}
        \bra{\alpha}
        \prod_{p=0}^{\Lambda-1} H_{a(p),b(p)} \ket{\alpha}},
    \end{split}
\end{equation}
\end{widetext}

where in the last step, we reexpressed the $q$ independent replicas in terms of a unique system at
inverse temperature $q \beta$. This is done by introducing a partition $(\Lambda_1,\dots, \Lambda_q)$ of the
operator string $S_\Lambda$ into $q$ parts such that the cutoffs sum up to $\Lambda$: $\sum_i \Lambda_i=\Lambda$. Clearly, in all the expressions above,
the numbers of (offdiagonal) operators $n_i$ ($n_2^{(i)}$) in slice $i$ of the operator string also
have to sum up to the complete number of (offdiagonal) operators $n$ ($n_2$). The introduction of the Kronecker delta
$\delta_{\alpha_{1,\Lambda_1},\dots,\alpha_{q,\Lambda_q}}$ expresses the fact that only operator strings in which
the states $\ket{\alpha_i,\Lambda_i}$ at the end of each slice $i$ are identical will contribute to our result.

Alltogether, we see that we have to perform an SSE calculation at inverse temperature $q\beta$ and
measure the observable:

\begin{equation}
   \delta_{\alpha_{1,\Lambda_1},\dots,\alpha_{q,\Lambda_q}}  \frac{ \Lambda! \prod_{i=1}^{q} (\Lambda_i-n_i)!}{ q^n (\Lambda-n)!
       \prod_{i=1}^q \Lambda_i!}.
\end{equation}

\section{R\'enyi entanglement entropies by $S_z$ sector}
\label{sec:szsector}
  Here, we provide some additional details on the reconstruction of the entanglement spectrum by
  symmetry sectors (here $S_z$ sectors) of the reduced density matrix using the method described in Refs.
  \onlinecite{song_bipartite_2012,chung_entanglement_2013}. Since the reduced density matrix $\rho_A$
  is block-diagonal with $S_z$ of subsystem $A$, we can split the calculation into the blocks
  $\rho_A^{S_z}$.

  The first observation is that the normalization of each block is no longer given by $1$, but by
  the probability $p_{S_z}$ of finding a state with subsystem magnetization equal to $S_z$:

  \begin{equation}
      \tr \rho_A^{S_z} = p_{S_z},
      \label{eq:trsz}
  \end{equation}
  which is easily measured in the independent ensemble. This is of course the first power sum of the
  eigenvalues of $\rho_A^{S_z}$.

  We can calculate the $q$-th power sum of $\rho_A^{S_z}$ by using the same method as described in
  the main text by just ignoring in the measurement states that are not in the correct sector. However, this would corresponds to a reduced density matrix
  $\tilde{\rho_A}$ which is normalized to $1$. In order to obtain the correct normalization, we
  calculate

  \begin{equation}
      \tr \left( \rho_A^{S_z}\right)^q = p_{S_z}^q \frac{p_{\varnothing\to
      A}(S_z)}{p_{A\to\varnothing}(S_z)},
  \end{equation}

  where $p_{\varnothing\to A}(S_z)$ and $p_{A \to \varnothing}(S_z)$ correspond to the transition
  probabilities estimated from measurements in the corresponding $S_z$ sector only. Note that these
  probabilities can be obtained by binning the measurements of the transition probabilities by their
  $S_z$ sectors \emph{without changing their normalization} and then by dividing by the probability
  of being in the correct sector in the corresponding ensemble (for the independent ensemble, this
  cancels exactly the factor $p_{S_z}^q$).

  From the knowledge of the power sums of the eigenvalues of $\rho_A^{S_z}$ we can now use the
  method described by Song {\it et al.}~\cite{song_bipartite_2012} to reconstruct the entanglement
  spectrum. Note however, that in equation (2.30) of Ref. \onlinecite{song_bipartite_2012} the $1$
  on the diagonal has to be replaced by the first power sum of the eigenvalues of $\rho_A^{S_z}$,
  {\it i.e.} by $p_{S_z} = \tr \rho_A^{S_z}$.

\bibliography{ee_sse}

\end{document}